\documentclass[letterpaper,twocolumn,10pt]{article}
\usepackage{usenix,epsfig,endnotes}
\usepackage{paralist}
\usepackage{listings}
\usepackage{commenting}
\usepackage{url}
\usepackage[breaklinks]{hyperref}
\usepackage{breakurl} 
\usepackage{cite}
\usepackage{color}
\usepackage{graphicx}
\usepackage{float}
\usepackage{caption}
\usepackage{textcomp}
\usepackage{xspace}
\usepackage{pifont}
\usepackage{subcaption}
\usepackage{multirow}
\usepackage{xcolor,colortbl}
\usepackage{authblk}
\usepackage{balance}

\declareauthor{bk}{Boris}{red!50}
\declareauthor{pv}{Pepe}{blue}

\begin{document}

\newcommand{\attacker}{\textit{A}\xspace}
\newcommand{\victim}{\textit{V}\xspace}
\newcommand{\provider}{\textit{T}\xspace}

\newcommand{\sender}{\textit{S}\xspace}
\newcommand{\receiver}{\textit{R}\xspace}

\newcommand{\paramP}{\textit{P}\xspace}
\newcommand{\paramWS}{\textit{windowSize}\xspace}
\newcommand{\paramWT}{\textit{windowType}\xspace}
\newcommand{\paramK}{k}
\newcommand{\paramF}{\textit{sampleFunction}\xspace}
\newcommand{\paramSP}{\textit{stepPattern}\xspace}
\newcommand{\paramL}{\textit{traceDuration}\xspace}

\newcommand{\monitorTool}{LoopScan\xspace}
\newcommand{\dtw}{d}
\newcommand{\approxdtw}{\hat{\dtw}}
\newcommand{\cmark}{\ding{51}}%
\newcommand{\xmark}{\ding{55}}%

\newcommand{\optionaltext}[1]{}

\lstdefinelanguage{JavaScript}{
  keywords={typeof, new, true, false, catch, function, return, null, catch, switch, var, if, in, while, do, else, case, break, let, const},
  keywordstyle=\color{blue}\bfseries,
  ndkeywords={class, export, boolean, throw, implements, import, this},
  ndkeywordstyle=\color{darkgray}\bfseries,
  identifierstyle=\color{black},
  sensitive=false,
  comment=[l]{//},
  morecomment=[s]{/*}{*/},
  commentstyle=\color{purple}\ttfamily,
  stringstyle=\color{red}\ttfamily,
  morestring=[b]',
  morestring=[b]"
}

\lstset{
   frame=tb,
   xleftmargin=.03\textwidth, xrightmargin=.01\textwidth,
   language=JavaScript,
   extendedchars=true,
   basicstyle=\footnotesize\ttfamily,
   showstringspaces=false,
   showspaces=false,
   numbers=left,
   numberstyle=\footnotesize,
   numbersep=9pt,
   tabsize=2,
   breaklines=true,
   showtabs=false,
   captionpos=b,
   upquote=true,
   belowskip=0pt
}

\definecolor{codegray}{gray}{0.95}
\newcommand{\code}[1]{\colorbox{codegray}{\texttt{#1}}}

\Urlmuskip=0mu plus 2mu

\date{}

\title{Loophole: Timing Attacks on Shared Event Loops in Chrome}

\author[*,$\dag$]{Pepe Vila}
\author[*]{Boris K{\"o}pf}

\affil[*]{IMDEA Software Institute}
\affil[$\dag$]{Technical University of Madrid (UPM)}
\affil[ ]{\textit {\{pepe.vila, boris.koepf\}@imdea.org}}

\maketitle


\subsection*{Abstract}
Event-driven programming (EDP) is the prevalent paradigm for graphical user
interfaces, web clients, and it is rapidly gaining importance for server-side
and network programming. Central components of EDP are {\em event loops},
which act as FIFO queues that are used by processes to store and dispatch
messages received from other processes.

In this paper we demonstrate that shared event loops are vulnerable to
side-channel attacks, where a spy process monitors the loop usage
pattern of other processes by enqueueing events and measuring the time
it takes for them to be dispatched. Specifically, we exhibit attacks
against the two central event loops in Google's Chrome web browser:
that of the I/O thread of the host process, which multiplexes all
network events and user actions, and that of the main thread of the
renderer processes, which handles rendering and Javascript tasks.

For each of these loops, we show how the usage pattern can be
monitored with high resolution and low overhead, and how this can be
abused for malicious purposes, such as web page identification, user
behavior detection, and covert communication.

\section{Introduction}\label{sec:intro} 
\bgroup
\let\thefootnote\relax\footnotetext{Original publication in the Proceedings of the 26th Annual USENIX Security Symposium (USENIX Security 2017).\\\url{https://www.usenix.org/conference/usenixsecurity17/technical-sessions/presentation/vila}}
\egroup

Event-driven programming (EDP) consists of defining responses to
events such as user actions, I/O signals, or messages from other
programs.  EDP is the prevalent programming paradigm for graphical
user interfaces, web clients, and it is rapidly gaining importance for
server-side and network programming.  For instance, the HTML5
standard~\cite{html5spec} mandates that user agents be implemented
using EDP, similarly, Node.js, memcached, and Nginx, also rely on EDP.

In EDP, each program has an {\em event loop} which consists of a FIFO queue and
a control process (or thread) that listens to events. Events that arrive are
pushed into the queue and are sequentially dispatched by the control process
according to a FIFO policy. A key feature of EDP is that high-latency (or
blocking) operations, such as database or network requests, can be handled
asynchronously: They appear in the queue only as events signaling start and
completion, whereas the blocking operation itself is handled elsewhere. In this
way EDP achieves the responsiveness and fine-grained concurrency required for
modern user interfaces and network servers, without burdening programmers with
explicit concurrency control.  
\begin{figure}[h]
    \includegraphics[width=0.9\columnwidth]{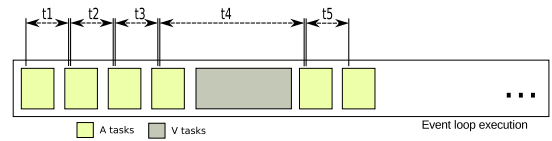}
    \caption{Shared event loop. \attacker enqueues multiple short
      tasks and records the time at which each of them is
      processed. The time difference between two consecutive tasks
      reveals whether \victim has posted tasks in-between, and how
      long they took to execute.}\label{fig:eventloopexec}
\end{figure}

In this paper we show that EDP-based systems are susceptible to side-channel
attacks. The key observation is that event loops form a resource that can be
shared between mutually distrusting programs.  Hence, contention of this
resource by one program can be observed by the others through variations in the
time the control process takes for dispatching their events.
Figure~\ref{fig:eventloopexec} illustrates such a scenario for a loop that is
shared between an attacker \attacker and a victim \victim.

Attacks based on observable contention of shared resources have a long
history~\cite{lampson1973note} and an active
present~\cite{bernstein2005cache,osvik2006cache,YaromF14}; however,
attacks against shared event loops have so far only been considered
from a theoretical point of view~\cite{KadloorKV16}. Here, we perform
the first attacks against real EDP-based systems. Specifically, we
target shared event loops in the two central processes of Google's
Chrome web browser: The {\em host process}, whose event loop is 
shared between all requests for common resources, such as network and
user interface; and the {\em renderer processes}, whose loops can be
shared between Javascript tasks of different tabs or iframes.

We build infrastructure that enables us to spy on both loops from a
malicious HTML page. This is facilitated by the asynchronous
programming model used in both Chrome and Javascript.  Asynchronous
function calls trigger new tasks that are appended to the same
queue, in contrast to synchronous calls which are simply pushed onto
the current task's call stack and executed without preemption,
blocking the loop.
\vspace{5px}
\begin{compactitem}
\item For the event loop of the renderer we rely on the
  \verb!postMessage! API, which is a Javascript feature for
  cross-window communication based on asynchronous callbacks. By
  posting messages to ourselves we can monitor the event loop with a
  resolution of $25\,\mu$s, with only one task in the loop at each
  point in time.
\item For the event loop of the host process we rely on two different
  mechanisms: network requests to non-routable IP addresses, which
  enter the loop and abort very quickly, providing a
  resolution of $500\,\mu$s; and SharedWorkers, whose messages pass
  through the event loop of the host process, providing a resolution
  of $100\,\mu$s.
\end{compactitem}
\vspace{5px}
We use the information obtained using these techniques in three different attacks:
\vspace{5px}
\begin{asparaenum}
\item We show how event delays during the loading phase, corresponding
  to resource requests, parsing, rendering and Javascript execution,
  can be used to uniquely identify a web page.
  Figure~\ref{fig:webfingerprinttraces} visualizes this effect
  using three representative web pages. While this attack shares the
  goal with the Memento attack~\cite{jana2012memento}, the channels
  are quite different: First, in contrast to Memento, we find that the
  relative ordering of events is necessary for successful
  classification, which motivates the use of dynamic time warping as a
  distance measure. Second, we show that page identification through
  the event loop requires only minimal training: we achieve
  recognition rates of up to $75\%$ and $23\%$ for the event loops of the
  renderer and host processes, respectively, for 500 main pages from
  Alexa's Top sites. These rates are obtained using only one sample of
  each page for the training phase.
\\
\item We illustrate how user actions in cross-origin pages can be
  detected based on the delays they introduce in the event
  loop. In particular, we mount an attack against
    Google OAuth login forms, in which we measure the time between
    keystrokes while the user is typing a password. The timing
    measurements we obtain from the event loop are significantly less
    noisy or require less privileges than from other
    channels~\cite{hogye2001analysis,PeepingTom09,GrussRS15}.
\begin{figure}[ht]
\includegraphics[width=\columnwidth]{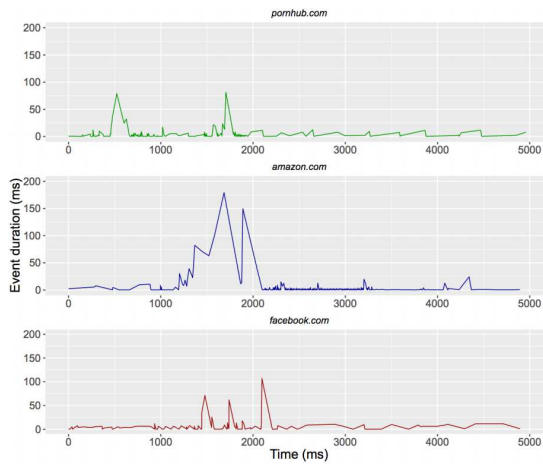}
\caption{Delays observed while loading different web pages, by an
  attacker tab sharing the renderer process.  Horizontal axis depicts
  elapsed real time, vertical axis depicts time taken by the event
  loop for processing the tasks inserted by the attacker. All pages
  are clearly distinguishable, both by the human eye and by
  classification techniques.}
\label{fig:webfingerprinttraces}
\end{figure}
\item We demonstrate that shared event loops can be used to transmit
information between cross-origin pages.
Specifically, we implement a covert channel with a bandwidth of
$200$\,bit/s through the renderer's main thread event loop, and
another one working cross-processes of $5$\,bit/s.
\end{asparaenum}

\vspace{5px}
Our attacks show that event loops can be successfully spied on even
with simple means. They work under the assumption that event loops
behave as FIFO queues; in reality, however, Chrome's event loop has a
more sophisticated structure, relying on multiple queues and a
policy-based scheduler. We believe that this structure
can be leveraged for much more powerful attacks in the future.

\section{Isolation Policies and Sharing of Event Loops in
  Chrome}\label{sec:background}

In this section we revisit the same origin policy and its variants.
We then discuss the relationship of these policies with the Chrome
architecture, where we put a special focus on the way in which event
loops are shared.

\subsection{Same Origin Policy} \label{ssec:background_sop}
The Same-Origin Policy (SOP) is a central concept in the web security
model: The policy restricts scripts on a web page to access data from
another page if their origins differ. Two pages have the same origin if
protocol, port and host are equal.

The demand for flexible cross-origin communication has triggered the
introduction of features such as domain relaxation, the postMessage
API, Cross-origin Resource Sharing (CORS), Channel Messaging,
Suborigins, or the Fetch API. This feature creep comes with an
increase in browser complexity and attack surface, which has motivated
browser vendors to move towards more robust multi-process
architectures.

\subsection{Overview of the Chrome Architecture} \label{ssec:background_arch}
The Chrome architecture is segmented into different operating system
processes. The rationale for this segmentation is twofold: to isolate
web content from the host~\cite{barth2008security}, and to support the
enforcement of origin policies by means of the
OS~\cite{reis2009isolating}. For achieving this segmentation, Chrome
relies on two processes: 

\begin{figure}[ht]
    \includegraphics[width=0.9\columnwidth]{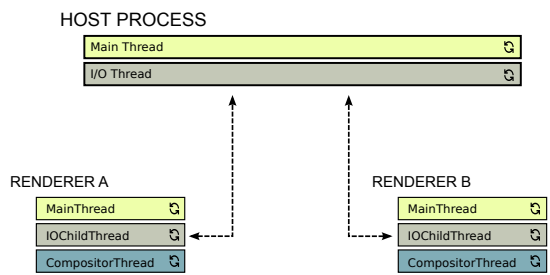}
    \caption{Overview of Chrome's architecture.}\label{fig:chromiumthreadsdiagram}
\end{figure}

The {\em host process} runs the top-level browser window. It has
access to system resources such as network, file system, UI events,
etc., which it manages on behalf of the unprivileged renderer
processes.  The host process runs several threads; the most relevant 
ones are:
\vspace{5px}
\begin{compactitem}
\item the \verb!CrBrowserMain! thread, which handles, e.g., user
  interaction events, and
\item the \verb!IOThread!, which handles, e.g., IPC, network stack,
  and file system.
\end{compactitem}
\vspace{5px}
\optionaltext{The host process also stores all cookies in
  a global \textit{cookie jar}. This makes stateful navigation across
  origins possible and provides the Internet's linkability
  capabilities, but it also introduces several privacy and security
  issues, see Section~\ref{sec:related}. To amend that, Chrome
  provides the \textit{Incognito} (or private) navigation mode, which
  enables a temporary and independent cookie jar.}

The {\em renderer processes} are sandboxed processes responsible for
parsing, rendering and Javascript
execution. Communication with the host process is done via an
inter-process communication (IPC) system based on message
passing. Each renderer runs several threads; the most relevant ones
are:
\vspace{5px}
\begin{compactitem}
\item the \verb!MainThread! where resource parsing, style calculation, layout,
    painting and non-worker Javascript runs, 
\item the \verb!IOChildThread!, which handles IPC communication with the host
    process, and
\item the \verb!CompositorThread!, which improves responsiveness during the
    rendering phase by allowing the user to scroll and see animations while the
    main thread is busy, thanks to a snapshot of the page's state.
\end{compactitem}
\vspace{5px}

Each of the threads in the host and renderer processes maintains at least
one event loop that is largely a FIFO queue. Inter-thread and
inter-process communication are carried out via message passing through
these queues. We next discuss scenarios where pages of different
origin can share the event loops of host and renderer processes. In
Section~\ref{sec:eavesdrop} we show how this sharing can be exploited
for eavesdropping.

\subsection{Sharing in the Renderer Processes}\label{ssec:share_rend}
Chrome supports different policies that govern how web applications
are mapped to renderer processes, and that influence whether or not
event loops are shared.

The default policy is called {\em process-per-site-instance}. It
requires using a dedicated renderer process for each instance of a
site. Here, a {\em site} is defined as a registered domain plus a
scheme. For example, https://docs.google.com and
https://mail.google.com:8080 are from the same site -- but not from
the same origin, as they differ in subdomain and port.  A {\em site
  instance} is a collection of pages from the same site that can
obtain references to each other (e.g., one page opened the other in a
new window using Javascript).

The other supported policies are more permissive. For example, the
{\em process-per-site} policy groups all instances of a site in the
same renderer process, trading robustness for a lower memory
overhead. The {\em process-per-tab} policy dedicates one renderer
process to each group of script-connected tabs. Finally, the {\em
single-process} policy lets both the host and renderer run
within a single OS process (only used for debugging purposes).

Even in the restrictive default process-per-site-instance policy,
there are some situations that force Chrome to host documents from
different sites in the same renderer process, causing them
to share the event loop:
\vspace{5px}
\begin{compactitem}
\item Iframes are currently hosted in the same process as their parent.
\item Renderer-initiated navigations such as link clicks, form submissions, and
    scripted redirections will reuse the same
    renderer as the origin page.
\item When the number of renderer processes exceeds a certain threshold,
Chrome starts to reuse existing renderers instead of creating new
ones.
\end{compactitem}
\vspace{5px}
On (64-bit) OSX and Linux, the threshold for reusing renderers is
calculated by splitting half of the physical RAM among the
renderers, under the assumption that each consumes 60MB.\footnote{On
  Android there is no threshold since the OS suspends idle processes.}
In our experiments, on a machine with $4$\,GB of RAM we could spawn 31
new tabs before any renderer was shared, whereas on a machine with
$8$\,GB of RAM we observed a threshold of approximately 70 renderers.
There is no apparent grouping policy for the pages that can share a
process when this threshold is exceeded, except for tabs in Incognito
mode not being mixed up with ``normal'' tabs. In particular, we do not
observe any preference for similar origins, same sites, or secure
versus insecure pages. In fact, even filesystem pages (loaded with
\verb!file://!) can co-reside with an arbitrary HTTP site.

\subsection{Sharing in the Host Process}\label{ssec:share_host}

The Chrome sandbox restricts access of renderers to privileged
actions. In particular, renderers have to communicate with the host
process for network requests or user input. The corresponding messages
of all renderers pass through the event loop of the host process' I/O
thread.

We illustrate this communication using two different examples: how
user actions flow from the host to the corresponding renderer process,
and conversely, how network requests flow from a renderer to the host
process.
%
\begin{compactitem}
\item {\em UI flow: } User actions such as mouse movements or clicks
  enter the browser through the main thread of the host process. The
  host main thread communicates the user event to the corresponding
  renderer by message passing between their I/O event loops, and the
  render acknowledges
  the receipt of this message. Even events with no Javascript
  listeners occupy the event loop of the renderer's main thread for a
  measurable interval.
\item {\em Net stack:} Chrome's net stack is a complex cross-platform
  network abstraction.  Any network request by a renderer is passed to
  the I/O thread of the host process, which forwards it to a global
  resource dispatcher that will pass it to a worker to fulfill the
  request. This worker will open a connection, if necessary, and
  request the resource. After the request is done, the response headers
  are received and sent back to the renderer process, which will respond
  with an ACK after reading, Finally, the body is received and the
  corresponding callbacks are triggered.
\end{compactitem}
\vspace{5px}

\section{Eavesdropping on Event Loops in Chrome}\label{sec:eavesdrop}

In this section we describe how to violate the SOP by eavesdropping on the
event loops of Chrome's host and renderer processes. For each of these
processes, we describe potential threat scenarios and present a simple
HTML page executing Javascript that can be used for spying. We then
present our monitoring tool to visualize the event loops of the browser.

\subsection{The Renderer Process Event Loop}\label{ssec:monitor_renderer}

\subsubsection{Threat Scenarios}
There are several scenarios in which an adversary site \attacker can
share the event loop of the renderer's main thread with a victim site
\victim. These scenarios are based on Chrome's policy for mapping
sites to renderers, see Section~\ref{ssec:share_rend}. We give two
examples:
  \begin{compactitem}
  \item \textit{Malicious advertisement}. In this scenario, \attacker
    runs as an advertisement iframed in \victim. The SOP protects
	\victim's privacy and itegrity by logically isolating both execution
	environments. However, \attacker's~iframe
    is able to execute Javascript on \victim's event loop, enabling it
    to gather information about the user behavior in \victim.
  \item \textit{Keylogger}. In this scenario, \attacker pops up a
    login form to authenticate its users via \victim's OAuth. Because
    the operation does not ask for special privileges and the password
	is never sent to \attacker, the victim could trust it and fill the form.
    Meanwhile, \attacker's page monitors keystroke timings (see
    Section~\ref{ssec:attacks_ui}), which can be used for recovering user
    passwords~\cite{sshkeystrokes2001}.
  \end{compactitem}

\subsubsection{Monitoring Techniques}\label{ssec:renderermonitor}
To monitor the renderer's event loop it is sufficient to continuously
post asynchronous tasks and measure the time interval between
subsequent pairs of events. We measure the monitoring resolution in
terms of the interval between two subsequent measurement events on an
otherwise empty loop.

The most common way of posting asynchronous tasks programmatically in
Javascript is \verb!setTimeout!. However, the resolution can be more
than $1000$\,ms for inactive tabs, rendering this approach useless for
the purpose of spying. To increase the resolution, we instead use the
\verb!postMessage! API for sending asynchronous messages to ourselves.

{\tt \small
\begin{figure}[h]
    \begin{lstlisting}[caption={Javascript code to monitor the main thread's event loop with the postMessage API.},label={lst:spy_rend}]
function loop() {
   save(performance.now())
   self.postMessage(0,'*')
} 
self.onmessage = loop
loop()
\end{lstlisting}
\end{figure}
}

The code in Listing~\ref{lst:spy_rend} shows how this is achieved.
The call to \verb!performance.now()! in line 2 of the function
\verb!loop! returns a high-resolution timestamp that is saved as
described below. The call to \verb!self.postmessage(0,'*')! in line 3
posts message ``0'' into the renderer's event loop, where the second
argument ``*'' indicates no restriction on the receiver's origin.
Line 5 registers the function \verb!loop! as an event listener, which enables it to
receive the messages it has posted. This causes \verb!loop! to
recursively post tasks, while keeping the render responsive since
other events are still being processed. 

In order to minimize the noise introduced by the measurement script
itself, the function~\verb!save! in line 2 uses a pre-allocated typed
array (\verb!Float64Array!) to store all the timing
measurements. Contrary to normal Javascript's sparse arrays, typed
arrays avoid memory reallocations and thus noisy garbage collection
rounds, see below. With that we achieve an average delay between two
consecutive tasks of around $25\,\mu$s on our target machine. This
resolution is sufficient to identify even short events. For example, a
single mouse movement event (without explicit event listener) consumes
around $100\,\mu$s. 

\subsubsection{Interferences}
In modern browsers there are several sources of noise that affect
measurement precision, beside the obvious effect of the underlying
hardware platform and OS. They include:
\begin{compactitem}
\item {\em Just-in-time compilation (JIT).} JIT can trigger code
  optimization or deoptimization, in the case of Chrome by the
  CrankShaft and Turbofan compilers, at points in time that are hard
  to predict. For our measurements we rely on a warm-up phase of
  about $150$\,ms to obtain fully optimized code.
\item {\em Garbage collection (GC).} In the case of V8, GC includes small
  collections (so-called {\em scavenges}) and major
  collections. Scavenges are periodical and fast ($<1$\,ms); but major
  collections may take $>100$\,ms, distributed into incremental steps.
  In our data, scavenges are easily identifiable due to their
  periodicity, while major collections could be spotted due to their characteristic
  size. On some browsers, such as Microsoft's Internet Explorer, GC rounds
  can be triggered programmatically, which helps to eliminate noise from the
  measurements enabling more precise attacks~\cite{dedupsp16}.
\vspace{5px}
\end{compactitem}
While all of these features reduce the effectiveness of our attacks,
it is interesting to think of them as potential side-channels by
themselves. For example, observable GC and JIT events can reveal
information about a program's memory and code usage patterns,
respectively~\cite{pedersentrash}.

\subsection{The Host Process Event Loop}\label{ssec:monitor_browser}

\subsubsection{Threat Scenarios}
The Chrome sandbox ensures that all of the renderer's network and user
interaction events pass through the host process' I/O event loop, see
Section~\ref{ssec:share_host}.  We describe two threat scenarios where
this could be exploited.
\begin{compactitem}
\item {\em Covert channel}. Pages of different origins running in
  different (disconnected) tabs can use the shared event loop to
  implement a covert channel, violating the browser's isolation
  mechanisms. This will work even if one (or both) pages run in
  incognito mode. This channel can be used for tracking users across
  sessions, or to exfiltrate information from suspicious web pages
  without network traffic.
  \item {\em Fingerprinting}. A tab running a rogue page of \attacker
    can identify which pages are being visited by the user in other
    tabs by spying on the shared event loop. Detecting the start of a
    navigation is facilitated by the fact that the I/O
    thread blocks for a moment when the user types in a URL and
    presses enter.
\end{compactitem}

\subsubsection{Monitoring Techniques}

There are many ways to post asynchronous tasks into the event loop of
the host process; they differ in terms of the resolution with which
they enable monitoring the event loop and the overhead they imply.
Below we describe two of the techniques we used.

\paragraph{Network Requests.} 

The first technique is to use network requests to systematically
monitor the event loop of the I/O thread of the host process. A valid
network request may take seconds to complete, with only the start and
end operations visible in the loop, which provides insufficient
resolution for monitoring.

To increase the resolution, we make use of {\em non-routable} IP
addresses. The corresponding requests enter the I/O thread's event
loop, are identified as invalid within the browser, and trigger the
callback without any DNS resolution or socket creation. This mechanism
provides a monitoring resolution of $500\, \mu$s and has the
additional benefit of being independent from network noise.

Listing~\ref{lst:spy_host} shows the code of our monitoring procedure.
We rely on the Javascript Fetch API for posting the network requests.
The Fetch API provides an interface for fetching resources using
{\em promises}, which are ideal to manage asynchronous computations
thanks to their simple syntax for handling callbacks.
In line 2 we request and save a high-resolution timestamp. In line 3
we request a non-routable IP address, and set the rejection callback of
the promise to self, to recursively run when the request fails.
{\tt \small
\begin{figure}[h]
\begin{lstlisting}[caption={Javascript code to monitor the host's I/O thread using network requests.},label={lst:spy_host}]
function loop() {
   save(performance.now())
   fetch(new Request('http://0/')).
      catch(loop)
} 
loop()
\end{lstlisting}
\end{figure}
}

\paragraph{Shared Workers.} The second technique relies on web workers,
which is a mechanism for executing Javascript in the background. Web
workers that are {\em shared} between multiple pages are usually implemented
in a dedicated OS process; this means they communicate via IPC and,
therefore, can be used to spy on the I/O thread of the host process. This
mechanism provides a monitoring resolution of $100\, \mu$s.
{\tt \small
\begin{figure}[h]
\begin{lstlisting}
onconnect = function reply(e) {
   let port = e.ports[0]
   port.onmessage = function() {
      port.postMessage(0)
   }
}
\end{lstlisting}
\begin{lstlisting}[caption={Javascript code to monitor the host's I/O
    thread using SharedWorkers. The first snippet is the worker's
    `pong.js' file. Second snippet is the Javascript code that monitors
    the I/O thread by communicating with the worker.},label={lst:spy_host2}]
const w = new SharedWorker('pong.js')
function loop() {
    save(performance.now())
    w.port.postMessage(0)
}
w.port.onmessage = loop
loop()
\end{lstlisting}
\end{figure}
}
Listing~\ref{lst:spy_host2} shows the code of our worker-based
monitoring procedure. The first snippet defines the worker's job,
which consists in replying to each received message. In the second
snippet, we register the worker in line 1. In lines 2-7 we record a
timestamp and recursively send messages to the worker, analogous to
Listing~\ref{lst:spy_rend}.  As a result, we measure the round-trip
time from the page to the worker, which reflects the congestion in the
I/O event loop. Note that one can further increase the
  measurement resolution by recording the time in each endpoint and
  merging the results.

\subsubsection{Interferences}
\label{ssec:crosspracticalchallenges}
There are many different sources of noise and uncertainty in the I/O thread of
the host process. The most notable ones include the interleaving with the host's
main thread and the messages from other renderers, but also the GPU process and
browser plugins. While these interferences could potentially be exploited as
side channels, the noise becomes quickly prohibitive as the loop gets crowded.

\subsection{The \monitorTool Tool}\label{ssec:monitor_tool}
We implement the eavesdropping techniques described in
Sections~\ref{ssec:monitor_renderer} and~\ref{ssec:monitor_browser} in
a tool called \monitorTool, which enables us to explore the
characteristics of the side channel caused by sharing event
loops. \monitorTool is based on a simple HTML page that monitors the
event loops of the host and renderer processes. It relies on the D3.js
framework, and provides interactive visualizations with minimap,
zooming, and scrolling capabilities, which facilitates the inspection
of traces. For example, Figure~\ref{fig:goauthkeystroke}
is based on a screenshot from \monitorTool.

\monitorTool's functionality is in principle covered by the
powerful Chrome Trace Event Profiling Tool (about:tracing)~\cite{traceEventProfiling},
which provides detailed flame graphs for all processes and
threads. However, \monitorTool has the advantage of delivering more
accurate timing information about event-delay traces than the profiler, since
loading a page with the Trace Event Profiling tool severely distorts the
measurements. \monitorTool source is publicly available at \url{https://github.com/cgvwzq/loopscan}.

\section{Attacks}\label{sec:attacks}

In this section we systematically analyze the side channel caused by
sharing event loops in three kinds of attacks: a page
identification attack, an attack where we eavesdrop on user actions,
and a covert channel attack. For all attacks we spy on the event loops
of the renderer and the host processes, as described in
Sections~\ref{ssec:monitor_renderer} and~\ref{ssec:monitor_browser}.
We performed these attacks over the course of
a year, always using the latest stable version of Chrome
(ranging from v52-v58). The results we obtain
are largely stable across the different versions.

\subsection{Page identification}\label{ssec:attacks_finger}

We describe how the event-delay trace obtained from spying on event
loops can be used for identifying webpages loaded in other tabs. We
begin by explaining our data selection and harvesting process and the
chosen analysis methods, then we describe our experimental setup and
the results we obtain.

\subsubsection{Sample Selection}
We start with the list of Alexa Top 1000 sites, from which we remove
duplicates. Here, duplicates are sites that share the subdomain but
not the top-level domains (e.g., ``google.br'' and ``google.com'') and
that are likely to have similar event-delay traces.  From the
remaining list, we randomly select 500 sites as our sample set.
This reduction facilitates a rigorous exploration of the data and
the parameter space.

\subsubsection{Data Harvesting}\label{ssec:dataharvesting}

We visit each page in the sample set 30 times for both the renderer
and the host process, to record traces of event-delays during the
loading phase.

The event-delay traces for the {\em renderer process} consist of
200.000 data items each. On our testing machine, the measurement
resolution (i.e. the delay between two subsequent measurement events
on an otherwise empty loop) lies at
approximately $25\,\mu$s. That is, each trace captures around 5
seconds (200.000$\cdot25\,\mu s=5$\,s) of the loading process of a
page in the sample set.

The event-delay traces for the {\em host process} consist of 100.000
data items each. The measurement resolution lies in the range of
$80-100\, \mu$s,
i.e. each trace captures around $9\,$s of the loading process of a
page.

We automate the harvesting procedure for the renderer process as
follows:
\begin{compactenum}
\item \label{l}{Open a new tab via\\ \verb!target = window.open(URL, '_blank');!~\footnote{Note that this requires disabling Chrome's
      popup blocker from ``chrome://settings/content''.}}
\item Monitor the event loop until the trace buffer is full
\item Close the tab
\item Send the trace to the server
\item Wait 5 seconds and go to \ref{l} with next URL
\end{compactenum}
The harvesting procedure for the host process differs only in that we
use the \verb!rel="noopener"! attribute in order to spawn a new
renderer.

We conducted measurements on the following three machines:
\begin{enumerate}
\item Debian
8.6 with kernel 3.16.0-4-amd64, running on an Intel~i5~@~3.30GHz~x~4
with 4~GB of RAM, and Chromium v53;
\item Debian 8.7 with
kernel 3.16.0-4-amd64, running on an Intel~i5-6500~@~3.20GHz~x~4 with
16~GB of RAM, and Chromium  v57; and
\item OSX running on a
Macbook~Pro~5.5 with Intel~Core~2~Duo~@~2.53GHz with 4~GB of RAM, and
Chrome v54.
\end{enumerate}

We measure the timing on a Chrome instance with two tabs, one for the
spy process and the other for the target page. For the renderer
process, we gather data on all machines; for the host process on
(2) and (3). Overall, we thus obtain five corpora of 15.000 traces 
each. 

\subsubsection{Classification}

\paragraph{Event Delay Histograms.}

Our first approach is to cluster the observed event delays around $k$
centers, and to transform each trace into a histogram that represents
the number of events that fall into each of the $k$ classes. We then
use the Euclidean distance as a similarity measure on the
$k$-dimensional signatures.

This approach is inspired by the notion of memprints
in~\cite{jana2012memento}. It appears to be suitable for classifying
event-delay traces obtained from event loops because, for example,
static pages with few external resources are more likely to produce
long events at the beginning and stabilize soon, whereas pages with
Javascript resources and animations are likely to lead to more irregular
patterns and produce a larger number of long delays.
Unfortunately, our experimental results were discouraging, with less
than a $15\%$ of recognition rate in small datasets.

\paragraph{Dynamic Time Warping.}

Our second approach is to maintain temporal information about the
observed events. However, the exact moments at which events occur are
prone to environmental noise. For example, network delays will
influence the duration of network requests and therefore the arrival
of events to the event loop. Instead, we focus on the
relative ordering of events as a more robust feature for page
identification.

This motivates the use of {\em dynamic time warping
  (DTW)}~\cite{berndt1994using} as a similarity measure on event-delay
traces. DTW is widely used for classifying time series, i.e.
sequences of data points taken at successive and equally spaced points
in time. DTW represents a notion of distance that considers as
``close'' time-dependent data of similar shape but different speed,
i.e. DTW is robust to horizontal compressions and stretches. This is
useful, for example, when one is willing to assign a low distance
score to the time series ``abc`` and ``abbbbc`, insensitive
to the prolonged duration of ``b``.
Formally, DTW compares two time series: a \textit{query},
$X = (x_{1},...,x_{n})$, and a \textit{reference},
$Y = (y_{1},...,y_{m})$. For that we use a non-negative distance
function $f(x_i,y_i)$ defined between any pair of elements $x_{i}$ and
$y_{j}$.
The goal of DTW is to find a matching of points in $X$ with points in
$Y$, such that (1) every point is matched, (2) the relative ordering
of points in each sequence is preserved (monotonicity), (3) and the cummulative
distance (i.e. the sum of the values of $f$) over all matching points
is minimized. This matching is called a {\em warping path}, and
the corresponding distance is the {\em time warping distance}
$\dtw(X,Y)$. 

\begin{figure}[h]
    \includegraphics[width=0.9\columnwidth]{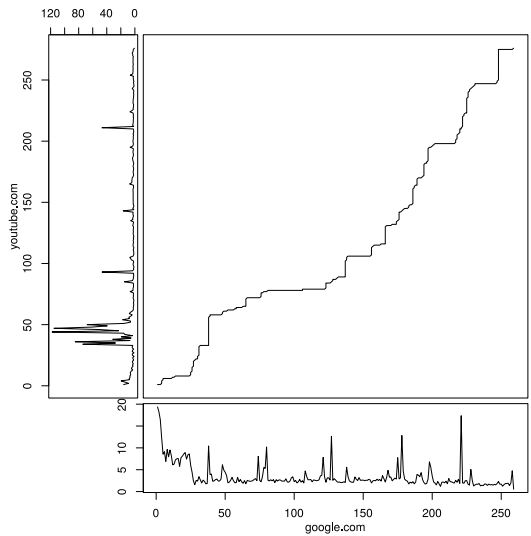}
    \caption{The path in the upper right square represents
the optimal alignment between points in the time series corresponding to 
'google.com' (horizontal axis) with points in the time series of
'youtube.com' (vertical axis).}
    \label{fig:warpingmatrix}
\end{figure}

Figure~\ref{fig:warpingmatrix} visualizes a warping path
between the time series corresponding to event-delay traces observed
while loading different webpages.

\subsubsection{Speed-up Techniques}\label{ssec:speedup}

\begin{figure}[ht]
    \includegraphics[width=0.9\columnwidth]{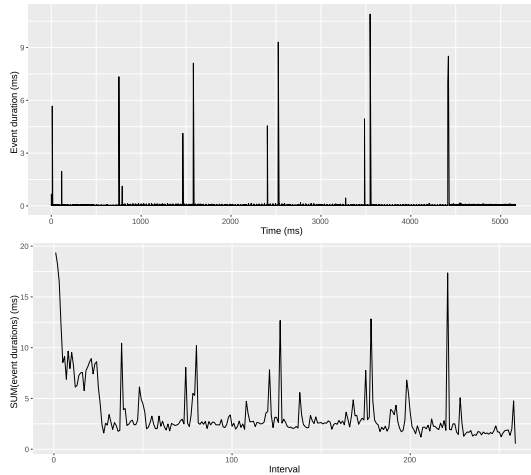}
    \caption{The top figure represents a raw trace of 200.000 time
      measurements from the renderer's main thread extracted while
      loading ``google.com''. The bottom figure displays the same data
      after being converted into a time series with $\paramP=20$\,ms,
      i.e. using only 250 data points. The difference
        in the height of the peaks is due to the accumulation of small
        events in the raw data, which are not perceptible in the top
        figure.}
    \label{fig:rawtracevsts}
\end{figure}

Unfortunately, the time required for computing $\dtw(X,Y)$ is
quadratic in the length of the input sequences and does not scale up
to the raw data obtained in our measurements. We rely on two kinds of
speed-up techniques, one at the level of the data and the other at the
level of the algorithm:

At the level of data, we reduce the dimension of our data by applying
a basic sampling algorithm: We split the raw trace into groups of
measurements corresponding to time intervals of duration $\paramP$, and
replace each of those groups by one representative. This
representative can be computed by summing over the group, or by taking
its average, maximum or minimum. The {\it sum} function generally yields
the best results among different sampling functions and is the one that
we use onwards. Sampling reduces the size of the traces by a factor of
$\paramP/t$, where $t$ is the average duration of an event delay.
Figure~\ref{fig:rawtracevsts} shows two plots with the raw data taken
from a renderer's main thread loop, and its corresponding time series
obtained after sampling.

At the algorithmic level, we use two sets of techniques for pruning
the search for the optimal warping path, namely windowing and
step patterns~\cite{JSSv031i07}.
\vspace{5px}
\begin{asparaitem}
\item {\em Windowing} is a heuristic that enforces a global constraint
  on the envelope of the warping path. It speeds up DTW but will not
  find optimal warping paths that lie outside of the envelope.  Two
  well-established constraint regions are the {\em Sakoe-Chiba band}
  and the {\em Itakura parallelogram}, see
  Figure~\ref{fig:windowtypes}.

\begin{figure}[ht]
    \includegraphics[width=0.9\columnwidth]{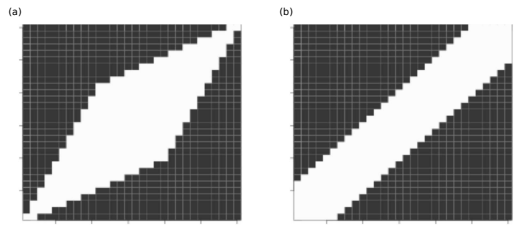}
    \caption{A global window constraint defines an envelope limiting
      the search space for optimal warping paths: (a) Itakura
      parallelogram, and (b) Sakoe-Chiba band.}
    \label{fig:windowtypes}
\end{figure}

\item {\em Step patterns} are a heuristic that puts a local constraint
  on the search
for a warping path, in terms of restrictions on its slope.  In
particular, we rely on three well-known step patterns available in R.
Intuitively, the {\em symmetric1} pattern favors progress
close to the diagonal, the {\em symmetric2} pattern allows for
arbitrary compressions and expansions, and the {\em
  asymmetric} forces each point in the reference to be used only
{\em once}. 
\end{asparaitem}

\begin{figure*}[ht]
    \includegraphics[width=0.9\textwidth]{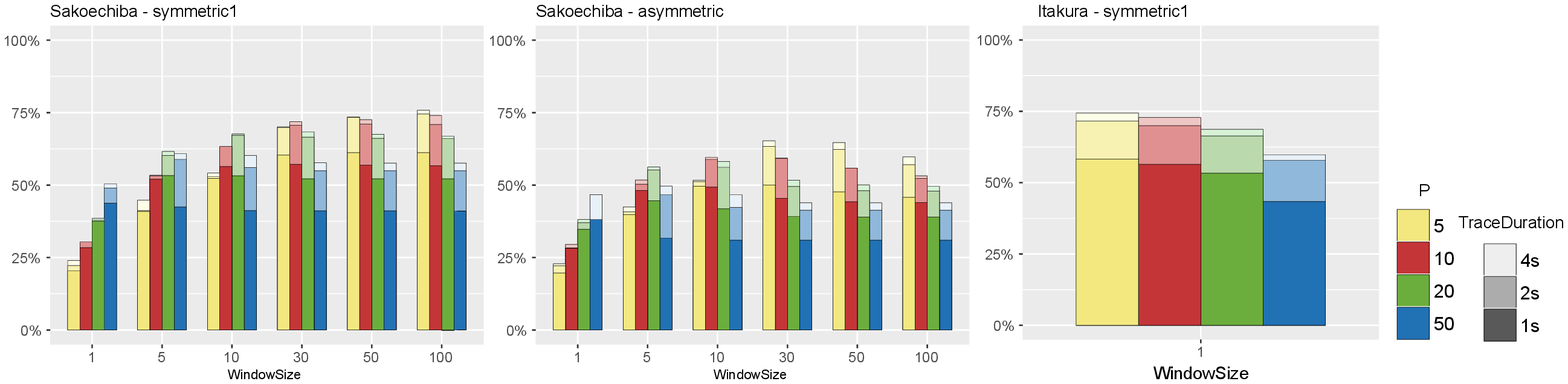}
    \caption{Web page identification performance after tuning with traces from the renderer on Linux machine (1). Effect of $\paramP$, $\paramL$, and $\paramWS$, with three combinations of $\paramSP$ and $\paramWT$.}
    \label{fig:tuningsummary}
\end{figure*}

\subsubsection{Parameter tuning}
 
The possible configurations of the techniques presented in
Section~\ref{ssec:speedup} create a large parameter
space, see Table~\ref{tab:parameterlist} for a summary.

\begin{table}[h]
\resizebox{\columnwidth}{!}{
    \begin{tabular}{|l|c|l|}
    \hline
    Parameter & Values & Description \\ 
    \hline
    $\paramL$ & $1000,2000,4000$ & Trace duration (ms) \\
    $\paramP$ & $5,10,20,50$ & Sampling interval (ms) \\
    $\paramWT$ & {\it itakura}, {\it sakoechiba} & Window constraint \\
    $\paramWS$ & $1,5,10,30,50,100$ & Window size \\
    $\paramSP$ & \begin{minipage}{0.5\columnwidth}
      {\it symmetric1}, {\it symmetric2},\newline \centering{\it asymmetric}
    \end{minipage}& Step pattern \\
    \hline
\end{tabular}
}
\caption{List of parameters tuned for optimizing web page identification}
\label{tab:parameterlist}
\end{table}

We systematically identify the optimal parameter configuration for
each event loop on each machine.  To avoid overfitting, we divide our
dataset of 30 traces (per page, loop, and machine) into 15 traces for
tuning and 15 for cross-validation. For each parameter configuration
we perform a lightweight version (with 3 rounds) of the evaluation
phase described in Section~\ref{sec:expres}. Figure~\ref{fig:tuningsummary}
visualizes an extract of the results we obtain for the renderer process
of the Linux (1) machine. The tuning phase yields the following insights:

\begin{asparaitem}
\item
The optimal parameters depend on the
  loop but appear to be stable across machines.
\item Measuring the loading phase during $2$ seconds is sufficient for
  recognition of a webpage; the gain in recognition from using longer
  traces is negligible.
\item $\paramP$ and $\paramWS$ are the parameters with the biggest
impact on the recognition rate. However, they also have the biggest
impact on the computational cost (the optimal choice being most expensive one).
\item The combination of $\paramSP=\textit{symmetric1}$ and
  $\paramWT=\textit{sakoechiba}$ generally yields the best results.
\end{asparaitem}

\subsubsection{Experimental Results}\label{sec:expres}

We evaluate the performance of page identification through the shared
event loops of host and renderer processes on each individual machine,
as well as through the renderer process across two different machines.

To this end, we select the top configuration for each corpus from the
tuning phase and carry out a 10-fold
cross-validation. In each of the 10 rounds, we
  partition the validation set into a training set that contains {\em
    one} trace of each page, and a testing set that contains {\em
    three} different (out of the 14 available) traces of each
  page. For each of the traces in the testing set, we compute the set
of $\paramK$ closest matches in the training set according to the time
warping distance.

We measure performance in terms of the {\em $\paramK$-match rate},
which is the percentage of pages in the testing set for which the true
match is within the set of $\paramK$ closest matches. We abbreviate
the $1$-match rate by {\em recognition rate}, i.e. the percentage of
pages where the best match is the correct one. The result of the
cross-validation is the average $\paramK$-match rate over all 10
rounds.

Table~\ref{tab:crossvalresults} summarizes our experiments. We
highlight the following results:
\begin{table}[h]
\resizebox{.95\columnwidth}{!}{
    \begin{tabular}{|l|l|c|c|c|c|}
        \cline{3-6}
        \multicolumn{2}{c|}{} & \multicolumn{4}{c|}{$k$} \\
        \cline{3-6}
        \multicolumn{2}{c|}{} & 1 & 3 & 5 & 10 \\
        \hline
        \multirow{2}{*}{\rotatebox[origin=c]{90}{(1)}}&\multirow{2}{*}{Renderer}& \cellcolor{blue!25}$76.7$\,\% & $86.7$\,\% & $88.8$\,\% & $91.1$\,\%
                \\
                \cline{3-6}
                & & \multicolumn{4}{c|}{$\textit{sym1,sakoe},\paramP=5,\paramWS=100$} \\
                \cline{3-6}
        \hline
        \multirow{4}{*}{\rotatebox[origin=c]{90}{(2)}}&\multirow{2}{*}{Renderer}& \cellcolor{blue!25}$58.2$\,\% & $68.6$\,\% & $71.8$\,\% & $75.1$\,\%
                \\
                \cline{3-6}
                & & \multicolumn{4}{c|}{$\textit{sym1,sakoe},\paramP=5,\paramWS=100$} \\
                \cline{3-6}
                & \multirow{2}{*}{I/O host} & \cellcolor{blue!25}$16.2$\,\% & $23.2$\,\% & $27.9$\,\% & $36.1$\,\% \\
                \cline{3-6}
                & & \multicolumn{4}{c|}{$\textit{sym1,sakoe},\paramP=20,\paramWS=30$} \\
        \hline
        \multirow{4}{*}{\rotatebox[origin=c]{90}{(3)}}&\multirow{2}{*}{Renderer}& \cellcolor{blue!25}$61.8$\,\% & $74.5$\,\% & $78.4$\,\% & $83.1$\,\%
                \\
                \cline{3-6}
                & & \multicolumn{4}{c|}{$\textit{sym1,sakoe},\paramP=5,\paramWS=100$} \\
                \cline{3-6}
                & \multirow{2}{*}{I/O host} & \cellcolor{blue!25}$23.48$\,\% & $32.9$\,\% & $38.1$\,\% & $46.6$\,\% \\
                \cline{3-6}
                & & \multicolumn{4}{c|}{$\textit{sym1,sakoe},\paramP=20,\paramWS=30$} \\
        \hline
    \end{tabular}
}
    \caption{10-fold cross-validation results on different machines and different event loops, with the best configuration after tuning.
	Machines (1) and (2) refer to the Linux desktops, (3) to the OSX laptop, as described in Section~\ref{ssec:dataharvesting}.}
\label{tab:crossvalresults}
\end{table}
\begin{asparaitem}
\item We can correctly identify a page by spying on the renderer from (1) in up to
   $76.7\%$ of the cases, and correctly narrow down to a set of 10
  candidates in up to $91.1\%$ of the cases.
\item We can correctly identify a page though the host process from (3) in
  up to $23.48\%$ of the cases, and narrow down to a set of 10 candidates in up to
  $46.6\%$ of the cases.
\item We stress that these recognition rates are obtained using
  a {\em single} trace for training.
\item Recognition is easier through the renderer than through the
  host. This is explained by the difference in noise and measurement
  resolution, see
  Section~\ref{ssec:crosspracticalchallenges}. Furthermore, most
  operations on the host only block the I/O thread while signaling
  their start and completion, whereas the renderer is blocked during
  the entire execution of each Javascript task.
\item We observe different recognition rates on different machines.
  However the homogeneity in hardware and software of Macbooks
  facilitate reuse of training data across machines, which may make
  remote page identification more feasible.
\item We obtain recognition rates below $5\,\%$ for recognition
  across machines (1) and (3).  A reason
  for this poor performance is that events on the OSX laptop often
  take 2x-5x more time than on the Linux desktop machine. This
  difference is reflected in the height of the peaks (rather
  than in their position), which is penalized by DTW. Normalizing the
  measurements could improve cross-machine recognition.
\end{asparaitem}

The code and datasets used for tuning and cross-validation are available
as an R library at \url{https://github.com/cgvwzq/rlang-loophole}.

\subsubsection{Threats to Validity}\label{ssec:crossthreatstovalidity}

We perform our experiments in a closed-world scenario with only 2 tabs
(the spy and the victim) sharing an event loop. In real world
scenarios there can be more pages concurrently running the browser,
which will make detection harder. The worst case
  for monitoring the host process occurs when a tab performs
  streaming, since the loop gets completely flooded. The renderer's
  loop, however, is in general more robust to noise caused by other
  tabs in the browser.

On the other hand, our attacks do not make any use of the pages'
source code or of details of Chrome's scheduling system with priority
queues, the GC with periodic scavenges, or the frame rendering tasks.
We believe that taking into account this information can significantly
improve an adversary's eavesdropping capabilities and enable attacks
even in noisy, open-world scenarios.

\subsection{Detecting User Behavior}\label{ssec:attacks_ui}
In this section we show that it is possible to detect user actions
performed in a cross-origin tab or iframe, when the renderer process
is shared. We first describe an attack recovering the inter-keystroke
timing information against Google's OAuth login forms, which provides
higher precision than existing network-based
attacks~\cite{sshkeystrokes2001}. 
\subsubsection{Inter-keystroke Timing Attack on Google's OAuth login form}\label{ssec:keystrokeattack}
 
Many web applications use the OAuth protocol for user
authentication. OAuth allows users to login using their identity with
trusted providers, such as Google, Facebook, Twitter, or Github. On
the browser, this process is commonly implemented as follows:
\begin{compactenum}
\item A web application \attacker pops up the login form of a trusted
  provider \provider;
\item User \victim types their (name and) password and submits the
  form to \provider;
\item \provider generates an authorization token.
\end{compactenum}

Because the window of the
login form shares the event loop with the opener's renderer, a
malicious \attacker can eavesdrop on the keystroke events issued by
the login form.

\begin{figure}[ht]
\includegraphics[width=\columnwidth]{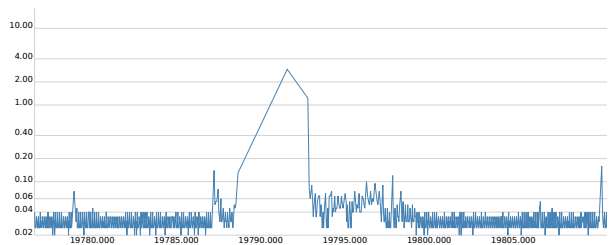}
\caption{Delay pattern generated by a keystroke in the Google OAuth
  login form, measured across origins on Chrome Canary v61 on OSX. The
  two consecutive delays of approx. $2$ms each, correspond to keydown and
  keypress event listeners.}
\label{fig:goauthkeystroke}
\end{figure}

Figure~\ref{fig:goauthkeystroke} depicts the event-delay trace of a
keystroke as seen by an eavesdropper on the renderer's event loop. The
trace contains two characteristic consecutive delays, caused by the
keydown and keypress event listeners. We use this observation to
identify keystrokes, by scanning the event-delay trace for pairs of
consecutive delays that are within a pre-defined range, forgoing any
training or offline work. Listing~\ref{lst:detect_keystroke} contains
the script that performs this operation. We define $0.4$\,ms as a lower
bound, and $3.0$\,ms as an upper bound for the range. We chose this
threshold before gathering the data, by manual inspection of a few
keystroke events. Note that this calibration could be done
automatically, based on the victim's interactions with a page
controlled by an attacker.

\begin{figure}[h]
  \begin{lstlisting}[caption={Pseudo-Javascript code to detect keystrokes in
      a trace of timestamps gathered by the code in
      Listing~\ref{lst:spy_rend}. We classify a timestamp as a
      keystroke if the differences to the previous and subsequent
      timestamps ($d1$ and $d2$) are both in a predefined
      range.},label={lst:detect_keystroke}]
const L = 0.4, U = 3.0, keys = []

for(let i=1; i<trace.length-1; i++){
    let d1 = trace[i] - trace[i-1],
        d2 = trace[i+1] - trace[i]

    if (L<d1<U && L<d1<U){
        keys.push(trace[i])
    }
}
\end{lstlisting}
\end{figure}

\subsubsection{Experimental Evaluation}

To evaluate the effectiveness of this attack, we have implemented a
malicious application \attacker that extracts the inter-keystroke timing
information from a user \victim logging-in via Google's OAuth. The focus of our
evaluation is to determine the accuracy with which keystroke timings
can be measured through the event loop. A full keystroke recovery
attack is out of scope of this paper; for this refer
to~\cite{sshkeystrokes2001}.

\begin{figure}[h]
    \includegraphics[width=0.95\columnwidth]{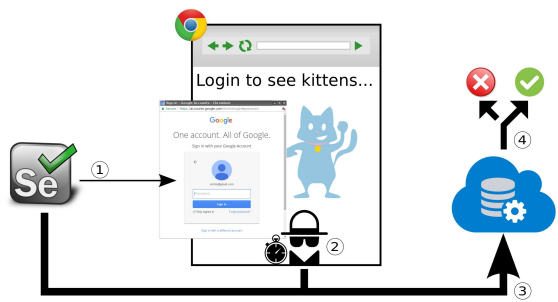}
    \caption{Experimental setup for evaluating effectiveness of
      automatic, cross-renderer keystroke detection.}
    \label{fig:seleniumDiagram}
\end{figure}

We simulate an inter-keystroke timing attack in 4 steps, which are
described below and illustrated in Figure~\ref{fig:seleniumDiagram}.
\begin{compactenum}
\item A Selenium\footnote{Selenium (\url{http://www.seleniumhq.org/}) is a
cross-platform testing framework for web applications that provides
capabilities for programmatically navigating to web pages and
producing user input.} script acting as~\victim navigates to \attacker,
  clicks on the login button (which pops up Google's OAuth login
  form), types a password, and submits the form. 
\item Meanwhile, the attacker~\attacker monitors the main thread's
  event loop using the attack described in
  Section~\ref{ssec:keystrokeattack}.
\item  \victim and \attacker send to the server the timestamps
  of the {\em real} and the {\em detected} keystrokes,
  respectively.
\item We compute the accuracy of the detected keystrokes, where we
  take the timestamps of the real keystrokes as ground truth.
  Matching the timestamps requires taking into account the delay
  ($6-12$~ms on our machine) between Selenium triggering an
  event, and Chrome receiving it.
\end{compactenum}

We use as inter-keystroke timings random delays uniformly drawn from
$100-300$\,ms. This choice is inspired by~\cite{hogye2001analysis}, who
report on an average inter-keystroke delay of $208$\,ms. Using random delays
is sufficient for evaluating the accuracy of eavesdropping on
keystrokes, but it obviously does not reveal any information about the
password besides its length.

\subsubsection{Experimental Results}

We perform experiments with 10.000 passwords extracted from
the RockYou dataset, where we obtain the following results:
\begin{itemize}
\item In $91.5$\% of the cases, our attack correctly identifies the
  length of a password.~\footnote{We configured Selenium to atomically
    inject characters that would require multiple keys to be pressed.}
  In $2.2$\% of the cases, the attack misses one or more characters,
  and in $6.3$\% of the cases it reports spurious characters.
\item For the passwords whose length was correctly identified, the
  average time difference between a true keystroke and a detected
  keystroke event is $6.3$ms, which we attribute mostly to the
  influence of Selenium.  This influence cancels out when we compute
  the average difference between a true inter-keystroke delay and a
  detected inter-keystroke delay, which amounts to $1.4$\,ms. The noise
  of these measurements is low: We observe a standard deviation of
  $6.1$\,ms, whereas the authors of~\cite{hogye2001analysis} report on
  $48.1$\,ms for their network based measurements.
\end{itemize}

Overall, our results demonstrate that shared event loops in Chrome
enable much
more precise recovery of keystroke timings than network-based
attacks. Moreover, this scenario facilitates to identify the time when
keystroke events enter the loop (from popping-up to form submission),
which is considered to be a major obstacle for inter-keystroke timing
attacks on network traffic~\cite{hogye2001analysis}. 

 Keystroke timing attacks based on monitoring
  \textit{procfs}~\cite{PeepingTom09} or CPU caches~\cite{GrussRS15}
  can extract more fine-grained information about keystrokes, such as
  containment in a specific subsets of keys. However, they require
  filesystem access or are more susceptible to noise, due to the
  resource being shared among all processes in the system. In
  contrast, our attack enables targeted eavesdropping without specific
  privileges.

\subsubsection{Open Challenges for Recognizing User Events}

\begin{figure}[h]
  \includegraphics[width=\columnwidth]{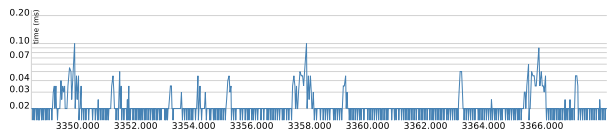}
  \caption{Mouse movement captured by~\monitorTool~tool. The graph
  shows 3 delays of $0.1$\,ms duration (at $t$ equals $3350$, $3358$ and $3366$),
  with an inter-event delay of $8$\,ms.}\label{fig:mousemovemonitor}
\end{figure}

We conclude by discussing two open challenges for re\-cognizing user
events, namely the detection of user events beyond keystrokes and the
detection of events in the browser's host process.
\paragraph{Detecting User Events beyond Keystrokes}
A continuous mouse movement results in a sequence of events, each of
which carrying information about the coordinates of the cursor's
trajectory. These events are issued with an inter-event delay of
$8$\,ms, and the (empty) event listener operation blocks the loop
for approx $~0.1$\,ms.  The particular frequency and duration of
these events makes mouse movements (or similar actions, like scrolling)
easy to spot with \monitorTool, as seen in Figure~\ref{fig:mousemovemonitor}.

Likewise, mouse click events, corresponding to ``up'' or ``down'', can
be identified using \monitorTool. Their shape depends on the specific
event listener of the spied web page and the HTML element being clicked.
We expect that events {\em with} specific listeners are more easily
detectable than events {\em without} registered event listeners, that is,
user actions that do not trigger Javascript execution. However, we can
use the context in which the event occurs to reduce the search space.
For instance, most mouse clicks only appear between two sequences of mouse
movement events.

We are currently investigating techniques that enable the automatic
identification of such patterns in event-delay streams.  A promising
starting point for this are existing on-line variants of dynamic
time-warping~\cite{StreamDTW07}.

\paragraph{Detecting User Events in the Host Process}
Our discussion so far has centered on detecting user events
in the event loop of the renderer process. However, all user events
originate in the main thread of the host process and are sent towards
a specific renderer through the event loop of the host's I/O
thread. Hence, any user action can in principle be detected by spying
on the host.

Unfortunately, our current methods are not precise enough for this task,
since the host's I/O thread is more noisy than the renderer's main thread
and the effect of a user action on the host process is limited to a short
signaling message, whereas the renderer's main thread is affected by the
execution of the corresponding Javascript event listener.

\subsection{Covert Channel}\label{ssec:attacks_covert}

In this section we show how shared event loops in Chrome can be abused
for implementing covert channels, i.e. channels for illicit
communication across origins. We first consider the case of
cross-origin pages sharing the event loop of a renderer's main thread
before we turn to the case of cross-origin pages sharing the event
loop of the host processes' I/O thread.

\subsubsection{Renderer Process}
We implement a communication channel to transmit messages from a
sender page \sender to a cross-origin receiver page \receiver running
in the same renderer process. 

For this, we use a simple, unidirectional transmission scheme without
error correction. Specifically, we encode each bit using a time
interval of fixed duration $t_b$. The optimal configuration of $t_b$
depends on the system. In our experiments we tried different values,
with $t_b=5$\,ms giving good results on different platforms:
Chromium~52.0 on Debian 64-bit and Chrome 53 on OSX.

In each of those intervals we do the following:
\begin{compactitem}
\item the sender \sender idles for transmitting a \verb!0!; it
executes a blocking task of duration $\hat{t} < t_b$ for transmitting a \verb!1!.
\item the receiver~\receiver monitors the event loop of the renderer's
  main thread using the techniques described in
  Section~\ref{ssec:monitor_renderer}; it decodes a \verb!0! if the
  length of the observed tasks is below a threshold (related to
  $\hat{t}$), and a \verb!1! otherwise.
\end{compactitem}
Transmission starts with \sender sending a \verb!1!, which is used by
the agents to synchronize their clocks and start counting time
intervals. Transmission ends with \sender sending a null byte.
With this basic scheme we achieve rates of $~200~bit/s$. These numbers can
likely be significantly improved by using more sophisticated coding schemes
with error correction mechanisms; here, we are only interested in the
proof-of-concept.

We note that there are a number of alternative covert channels for
transmitting information between pages running in the same
renderer~\cite{sopCovertChannels},
e.g.,
using \verb!window.name!, \verb!location.hash!, \verb!history.length!,
scrollbar's position or \verb!window.frames.length!. What
distinguishes the event-loop based channel is that it does not require
the sender and receiver to be connected, i.e. they do not need to hold
references to each other in order to communicate.

\subsubsection{Host Process}
\begin{figure}[h]
\includegraphics[width=.95\columnwidth]{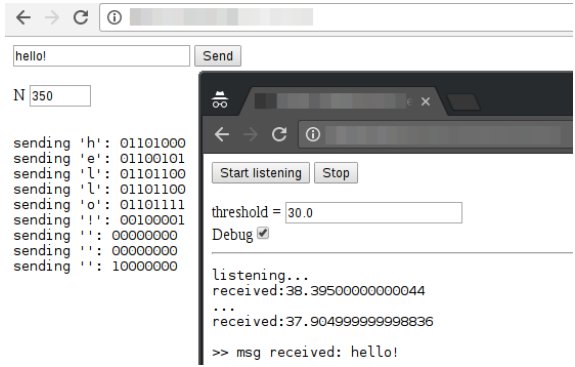}
    \caption{Covert channel through the I/O event loop of the Chrome's host process.
    Tabs in different renderer processes (one of them navigating in
    \textit{Incognito} mode) communicate.}
\label{fig:covertchannel}
\end{figure}

We also implement a communication channel to transmit messages between
two cooperative renderer processes sharing the host process. 
Transmission is unidirectional from sender \sender to receiver
\receiver. Figure \ref{fig:covertchannel} visualizes how this channel
can be used, even if one of the parties browses in Incognito mode.

As before, we encode each bit using a time intervals of fixed duration
$t_b$. During each intervals we do the following:
\begin{compactitem}
    \item the sender \sender idles for transmitting a \verb!0!; it posts $N$ fetch
        requests into the I/O thread's queue for sending a \verb!1!.
    \item the receiver \receiver monitors the event loop of the I/O
      thread of the host process using the techniques described in
      Section~\ref{ssec:monitor_browser}.
      It decodes a \verb!0! if the number of observed events during
      time interval $t_b$ is below a threshold, and \verb!1! otherwise.
\end{compactitem}
The optimal values of $N$ and $t_b$ highly depend on the machine.  In
our experiments we achieve good results, working on different systems,
with a $t_b = 200$\,ms and $N=350$, which give us a $5$~bit/s
transmission rate.  This rate is significantly lower than for
communication using the renderer event loop, which is explained by the
difference in noise and monitoring resolution of both channels, as
discussed in Section~\ref{ssec:crosspracticalchallenges}.

The threat scenario of this covert channel is more relevant than the
previous one for the renderer loop. For example it could be used for
exfiltrating information from an attacked domain (on a tab executing
malicious Javascript). Using Workers (which are
background threads that run independently of the user interface) we
can transfer information across origins, without affecting the user
experience and without generating network traffic.

\section{Discussion}\label{sec:discussion}

We have shown how sharing event loops leads to timing side-channels
and presented different attacks on Chrome.  We
  communicated our findings to the Chromium security team, who decided
  not to take action for the time being. Nevertheless, our results
  point to fundamental security issues in the event-driven
  architecture of browsers that eventually need to be addressed in a
  fundamental manner.
Below, we discuss how other platforms are affected and present
possible countermeasures.

\subsection{Beyond Chrome}
We focus on Chrome in our analysis because it is the most widely used
browser, and because it was the first one to implement a multi-process
architecture. However, there are good reasons to expect similar side
channels in other browsers, as they all follow the same event-driven
paradigm and rely on similar architectures. 

For instance, recent Firefox versions with
  multi-process support\footnote{Firefox's Electrolysis (or e10s)
    project} also rely on a privileged {\em browser process} and
  multiple {\em content processes} that, unlike renderers in Chrome,
  act as a pool of threads for each different origin (each with its
  own message queue).  Despite this difference, tests with
  \monitorTool on Firefox version 55 show that congestion on both
  event loops is observable across origins and tabs.

Specifically, we applied the monitoring technique for
  the renderers described in Section~\ref{ssec:renderermonitor} on a
  micro-benchmark with a set of 30 pages with 15 traces each. We 
  achieved a recognition rate of $49\%$, which is below the
  recognition rate achieved on Chrome for a set of 500 pages.  A fair
  comparison between both architectures will require a better
  understanding of Firefox's policy for mapping sites to threads and
  events to loops.
\subsection{Countermeasures}\label{ssec:countermeasures}

The attacks presented in this paper rely on two capabilities of
the adversary: (1) the ability to post tasks into the loop's queue
with high frequency, and (2) the ability to accurately measure the
corresponding time differences.

\paragraph{ Rate Limiting.}
An obvious approach to counter (1) is to impose a limit on the rate at
which tasks can be posted into an event loop. Unfortunately, rate
limiting implies penalties on performance, which is especially
proble\-matic for asynchronous code.

At the level of the renderer, one possibility is to rely on
an~\textit{accumulate and serve} policy~\cite{KadloorKV16}. With this
policy, the event loop accumulates all the incoming jobs in a buffer
for a period $T$, and then process and serves all the accumulated
jobs from party \attacker, followed by all the jobs from \victim.
This has the advantage of limiting the amount of
information leaked while retaining high amortized throughput.

At the level of the host process, where resource fetching is one of the
main performance concerns, setting any bound on the processing rate is
not acceptable. Here, it seems more reasonable to monitor the IPC
activity of all renderers and penalize or flag those who exhibit a
bad or anomalous behavior, e.g., along the lines
of~\cite{zhang2016cloudradar}.

\paragraph{Reduce Clock Resolution.}
An obvious approach to counter (2) is to limit the resolution of
available clocks. This has already been applied by browser vendors for
mitigating other kinds timing channels, but these efforts are unlikely
to succeed, as shown in~\cite{Browsertiming16}: Modern browsers have a
considerable number of methods to measure time without any explicit
clock. For instance, some recent exploits~\cite{gras_aslr_2017} use high-resolution
timers build on top of SharedArrayBuffers.
The current resolution of \verb!performance.now! is limited to $5\,\mu$s,
which makes microarchitectural timing attacks difficult, but does not
preclude the detection of Javascript events.

\paragraph{Full Isolation.}
As discussed in Section~\ref{ssec:background_arch}, Chrome's multi-process
architecture tries to use a different renderer for different origins,
except for some corner cases. The ``Site Isolation Project''
is an ongoing effort to ensure a complete process-per-site-instance
policy, that means: providing cross-process navigations, cross-process
Javascript interactions and out-of-process iframes. All this without
inducing too much overhead.

One open question is how to handle the system's process limit,
namely which sites should have isolation preference, or which
heuristic for process reuse should be used. A recent
proposal, ``IsolateMe''~\cite{isolationexplainer16}, puts the
developers in charge of requesting to be isolated from other web
content (even if it does not provide a firm guarantee).

\paragraph{CPU Throttling.}
 Chrome~v55 introduces an API that allows to limit how
  much CPU a background page is allowed to use, and to throttle tasks
  when they exceed this limit. This affects background tabs trying to
  spy on the renderer's main thread, but still allows spying on (and
  from) any iframe and popup, as well as on the I/O thread of the host
  process through shared Workers. Moreover, background tabs with audio
  activity are not affected, as they are always marked as foreground.
  Since Chrome~v57 pages (or tabs) are only subjected to throttling
  after 10 seconds in the background, which is too long to prevent the
  attacks in this paper.

\section{Related Work}\label{sec:related}

Timing attacks on web browsers date back to Felten and
Schneider~\cite{FeltenS00webtiming}, who use the browser cache to
obtain information about a user's browsing history.

More recently, so-called cross-site timing
attacks~\cite{Bortz07webtiming,VanGoethem15webtiming} have exploited
the fact that the browser attaches cookies to all requests, even when
they are performed across origins. The presence or absence of these
cookies can be determined by timing measurements, which reveals
information about the user's state on arbitrary sites. A special case
are cross-site search attacks~\cite{XSattack15}, which circumvent the
same-origin policy to extract sensitive information, by measuring the
time it takes for the browser to receive responses to search queries.

Other classes of browser-based timing attacks exploit timing
differences in rendering
operations~\cite{Kotcher13pixel,Stone13pixelperfect,AndryscoKMJLS15},
or simply use the browser as an entry point for Javascript that
exploits timing channels of underlying hardware, for example
caches~\cite{Oren2015spysandbox,gras_aslr_2017}, DRAM buffers~\cite{GrussMM16}, or CPU
contention~\cite{notSoIncognito15}.

Of those approaches, ~\cite{notSoIncognito15} is related to our work
in that it identifies web pages across browser tabs, based on timing
of Javascript and a classifier using dynamic time warping. However,
because the attack relies on CPU contention as a channel, it requires
putting heavy load on all cores for monitoring.  In contrast, our
attack exploits the browser's event loop as a channel, which can be
monitored by enqueing one event at a time. This makes our attack 
stealthy and more independent of the execution platform.

To the best of our knowledge, we are first to mount side-channel
attacks that exploit the event-driven architecture of web
browsers. Our work is inspired by a proof-of-concept
attack~\cite{matryoshka13} that steals a secret from a cross-origin
web application by using the single-threadedness of Javascript.  We
identify Chrome's event-driven architecture as the root cause of this
attack, and we show how this observation generalizes, in three
different attacks against two different event loops in Chrome.

Finally, a central difference between classical site fingerprinting~\cite{PanchenkoLPEZHW16,robustFingerprintUS16,identificationHttps02,peakaboo12}
approaches and our page identification attack is the adversary model:
First, our adversary only requires its page to be
opened in the victim's browser. Second, instead of traffic patterns in
the victim's network, our adversary observes only time delays in the
event queues of the victim's browser. We believe that our preliminary
results, with up to $76\%$ of recognition rate using {\em one single} sample
for training in a closed-world with 500 pages, can be significantly
improved by developing domain-specific classification techniques.

\section{Conclusions}

In this paper we demonstrate that shared event loops in Chrome are
vulnerable to side-channel attacks, where a spy process monitors the
loop usage pattern of other processes by enqueueing tasks and
measuring the time it takes for them to be dispatched.  We
systematically study how this channel can be used for different
purposes, such as web page identification, user behavior detection,
and covert communication.

\paragraph{Acknowledgments}
We thank Thorsten Holz, Andreas Rossberg, Carmela Troncoso, and the
anonymous reviewers for their helpful comments. We thank Javier Prieto
for his help with the data analysis. This work was supported by Ram{\'o}n y
Cajal grant RYC-2014-16766, Spanish projects TIN2012-39391-C04-01
StrongSoft and TIN2015-70713-R DEDETIS, and Madrid regional project
S2013/ICE-2731 N-GREENS.

\balance

{\footnotesize \bibliographystyle{acm}
\bibliography{websc}}

\begin{thebibliography}{10}

\bibitem{sopCovertChannels}
Covert channels in the sop.
\newblock \url{https://github.com/cgvwzq/sop-covert-channels}.
\newblock Accessed: 2017-02-16.

\bibitem{html5spec}
{HTML Living Standard}.
\newblock \url{https://html.spec.whatwg.org/}.
\newblock Accessed: 2017-05-24.

\bibitem{traceEventProfiling}
Understanding about:tracing results.
\newblock
  \url{https://www.chromium.org/developers/how-tos/trace-event-profiling-tool/trace-event-reading}.
\newblock Accessed: 2017-02-16.

\bibitem{isolationexplainer16}
Isolation explainer.
\newblock \url{https://wicg.github.io/isolation/explainer.html}, 2016.
\newblock Accessed: 2017-05-24.

\bibitem{AndryscoKMJLS15}
{\sc Andrysco, M., Kohlbrenner, D., Mowery, K., Jhala, R., Lerner, S., and
  Shacham, H.}
\newblock On subnormal floating point and abnormal timing.
\newblock In {\em SSP\/} (2015), IEEE.

\bibitem{barth2008security}
{\sc Barth, A., Jackson, C., Reis, C., Team, T., et~al.}
\newblock The security architecture of the chromium browser.
\newblock \url{http://www.adambarth.com/papers/2008/barthjackson-reis.pdf},
  2008.

\bibitem{berndt1994using}
{\sc Berndt, D.~J., and Clifford, J.}
\newblock Using dynamic time warping to find patterns in time series.
\newblock In {\em KDD workshop\/} (1994), {AAAI} Press.

\bibitem{bernstein2005cache}
{\sc Bernstein, D.}
\newblock {Cache-timing attacks on AES}.
\newblock \url{https://cr.yp.to/antiforgery/cachetiming-20050414.pdf}, 2005.

\bibitem{notSoIncognito15}
{\sc Booth, J.~M.}
\newblock Not so incognito: Exploiting resource-based side channels in
  javascript engines.
\newblock \url{http://nrs.harvard.edu/urn-3:HUL.InstRepos:17417578}, 2015.

\bibitem{Bortz07webtiming}
{\sc Bortz, A., and Boneh, D.}
\newblock Exposing private information by timing web applications.
\newblock In {\em WWW\/} (2007), ACM.

\bibitem{dedupsp16}
{\sc Bosman, E., Razavi, K., Bos, H., and Giuffrida, C.}
\newblock {Dedup Est Machina: Memory Deduplication as an Advanced Exploitation
  Vector}.
\newblock In {\em SSP\/} (2016), IEEE.

\bibitem{peakaboo12}
{\sc Dyer, K.~P., Coull, S.~E., Ristenpart, T., and Shrimpton, T.}
\newblock {Peek-a-Boo, I Still See You: Why Efficient Traffic Analysis
  Countermeasures Fail}.
\newblock In {\em SSP\/} (2012), IEEE.

\bibitem{FeltenS00webtiming}
{\sc Felten, E.~W., and Schneider, M.~A.}
\newblock Timing attacks on web privacy.
\newblock In {\em CCS\/} (2000), ACM.

\bibitem{XSattack15}
{\sc Gelernter, N., and Herzberg, A.}
\newblock {Cross-Site Search Attacks}.
\newblock In {\em CCS\/} (2015), ACM.

\bibitem{JSSv031i07}
{\sc Giorgino, T.}
\newblock Computing and visualizing dynamic time warping alignments in r: The
  dtw package.
\newblock {\em JSS 31}, 7 (2009), 1--24.

\bibitem{gras_aslr_2017}
{\sc Gras, B., Razavi, K., Bosman, E., Bos, H., and Giuffrida, C.}
\newblock {ASLR} on the {Line}: {Practical} {Cache} {Attacks} on the {MMU}.
\newblock In {\em {NDSS}\/} (2017), The Internet Society.

\bibitem{GrussMM16}
{\sc Gruss, D., Maurice, C., and Mangard, S.}
\newblock Rowhammer.js: {A} remote software-induced fault attack in javascript.
\newblock In {\em DIMVA\/} (2016), Springer.

\bibitem{GrussRS15}
{\sc Gruss, D., Spreitzer, R., and Mangard, S.}
\newblock Cache template attacks: Automating attacks on inclusive last-level
  caches.
\newblock In {\em USENIX Security\/} (2015), USENIX Association.

\bibitem{robustFingerprintUS16}
{\sc Hayes, J., and Danezis, G.}
\newblock {k-fingerprinting: A Robust Scalable Website Fingerprinting
  Technique}.
\newblock In {\em USENIX Security\/} (2016), USENIX Association.

\bibitem{hogye2001analysis}
{\sc Hogye, M.~A., Hughes, C.~T., Sarfaty, J.~M., and Wolf, J.~D.}
\newblock Analysis of the feasibility of keystroke timing attacks over ssh
  connections.
\newblock
  \url{http://www.cs.virginia.edu/~evans/cs588-fall2001/projects/reports/team4.pdf},
  2001.

\bibitem{jana2012memento}
{\sc Jana, S., and Shmatikov, V.}
\newblock {Memento: Learning secrets from process footprints}.
\newblock In {\em SSP\/} (2012), IEEE.

\bibitem{KadloorKV16}
{\sc Kadloor, S., Kiyavash, N., and Venkitasubramaniam, P.}
\newblock Mitigating timing side channel in shared schedulers.
\newblock {\em {IEEE/ACM} Trans. Netw. 24}, 3 (2016), 1562--1573.

\bibitem{Browsertiming16}
{\sc Kohlbrenner, D., and Shacham, H.}
\newblock {Trusted Browsers for Uncertain Times}.
\newblock In {\em USENIX Security\/} (2016), USENIX Association.

\bibitem{Kotcher13pixel}
{\sc Kotcher, R., Pei, Y., Jumde, P., and Jackson, C.}
\newblock {Cross-origin pixel stealing: timing attacks using CSS filters}.
\newblock In {\em CCS\/} (2013), ACM.

\bibitem{lampson1973note}
{\sc Lampson, B.~W.}
\newblock A note on the confinement problem.
\newblock {\em Communications of the ACM 16}, 10 (1973), 613--615.

\bibitem{Oren2015spysandbox}
{\sc Oren, Y., Kemerlis, V.~P., Sethumadhavan, S., and Keromytis, A.~D.}
\newblock {The Spy in the Sandbox: Practical Cache Attacks in JavaScript and
  Their Implications}.
\newblock In {\em CCS\/} (2015), ACM.

\bibitem{osvik2006cache}
{\sc Osvik, D.~A., Shamir, A., and Tromer, E.}
\newblock {Cache attacks and countermeasures: the case of AES}.
\newblock In {\em CT-RSA\/} (2006), Springer.

\bibitem{PanchenkoLPEZHW16}
{\sc Panchenko, A., Lanze, F., Pennekamp, J., Engel, T., Zinnen, A., Henze, M.,
  and Wehrle, K.}
\newblock Website fingerprinting at internet scale.
\newblock In {\em NDSS\/} (2016), The Internet Society.

\bibitem{pedersentrash}
{\sc Pedersen, M.~V., and Askarov, A.}
\newblock {From Trash to Treasure: Timing-sensitive Garbage Collection}.
\newblock In {\em SSP\/} (2017), IEEE.

\bibitem{reis2009isolating}
{\sc Reis, C., and Gribble, S.~D.}
\newblock Isolating web programs in modern browser architectures.
\newblock In {\em EuroSys\/} (2009), ACM.

\bibitem{StreamDTW07}
{\sc Sakurai, Y., Faloutsos, C., and Yamamuro, M.}
\newblock Stream monitoring under the time warping distance.
\newblock In {\em ICDE\/} (2007), IEEE.

\bibitem{sshkeystrokes2001}
{\sc Song, D.~X., Wagner, D., and Tian, X.}
\newblock {Timing Analysis of Keystrokes and Timing Attacks on SSH}.
\newblock In {\em USENIX Security\/} (2001), USENIX Association.

\bibitem{Stone13pixelperfect}
{\sc Stone, P.}
\newblock Pixel perfect timing attacks with html5 (white paper).
\newblock
  \url{https://www.contextis.com/documents/2/Browser_Timing_Attacks.pdf}, 2013.

\bibitem{identificationHttps02}
{\sc Sun, Q., Simon, D.~R., Wang, Y.-M., Russell, W., Padmanabhan, V.~N., and
  Qiu, L.}
\newblock Statistical identification of encrypted web browsing traffic.
\newblock In {\em SSP\/} (2002), IEEE.

\bibitem{VanGoethem15webtiming}
{\sc Van~Goethem, T., Joosen, W., and Nikiforakis, N.}
\newblock {The Clock is Still Ticking: Timing Attacks in the Modern Web}.
\newblock In {\em CCS\/} (2015), ACM.

\bibitem{matryoshka13}
{\sc Vela, E.}
\newblock Matryoshka: Timing attacks against javascript applications in
  browsers.
\newblock
  \url{http://sirdarckcat.blogspot.com.es/2014/05/matryoshka-web-application-timing.html},
  2013.

\bibitem{YaromF14}
{\sc Yarom, Y., and Falkner, K.}
\newblock {FLUSH+RELOAD:} {A} high resolution, low noise, {L3} cache
  side-channel attack.
\newblock In {\em {USENIX} Security Symposium\/} (2014).

\bibitem{PeepingTom09}
{\sc Zhang, K., and Wang, X.}
\newblock Peeping tom in the neighborhood: Keystroke eavesdropping on
  multi-user systems.
\newblock In {\em USENIX Security\/} (2009), USENIX Association.

\bibitem{zhang2016cloudradar}
{\sc Zhang, T., Zhang, Y., and Lee, R.~B.}
\newblock {CloudRadar: A Real-Time Side-Channel Attack Detection System in
  Clouds}.
\newblock In {\em RAID\/} (2016), Springer.

\end{thebibliography}

\end{document}